\documentclass[a4paper,11pt]{article}
\pdfoutput=1 

\usepackage{jheppub}

\usepackage{amsmath,amssymb,amsfonts,physics}
\usepackage{mathrsfs}
\usepackage{bbm}
\usepackage{slashed}
\usepackage{graphicx}
\usepackage{verbatim}
\usepackage[T1]{fontenc}
\usepackage[utf8]{inputenc}
\usepackage{mathbbol}
\usepackage[dvipsnames]{xcolor}
\usepackage[normalem]{ulem}

\definecolor{jbgreen}{rgb}{0.0, 0.65, 0.31}

\usepackage[backend=bibtex, sorting=none]{biblatex}
\bibliography{references}

\title{Scale-Invariance at the Core of Quantum Black Holes}

\author[a,b]{Johanna N.~Borissova,}
\author[c,d]{Aaron Held,}
\author[a,b,e]{Niayesh Afshordi}

\affiliation[a]{Perimeter Institute for Theoretical Physics, 31 Caroline Street North, Waterloo, ON, N2L 2Y5, Canada}
\affiliation[b]{Department of Physics and Astronomy,  University of Waterloo, 200 University Avenue West, Waterloo, ON, N2L 3G1, Canada}
\affiliation[c]{
Theoretisch-Physikalisches Institut, Friedrich-Schiller-Universit\"at Jena,
Max-Wien-Platz 1, 07743 Jena, Germany}

\affiliation[d]{
The Princeton Gravity Initiative, Jadwin Hall, Princeton University,
Princeton, New Jersey 08544, U.S.A.}
\affiliation[e]{Waterloo Centre for Astrophysics, University of Waterloo, Waterloo, ON, N2L 3G1, Canada}

\emailAdd{jborissova@pitp.ca}
\emailAdd{aaron.held@uni-jena.de}
\emailAdd{nafshordi@pitp.ca}

\abstract{We study spherically-symmetric solutions to a modified Einstein-Hilbert action with Renormalization Group scale-dependent couplings, inspired by Weinberg's Asymptotic Safety scenario for Quantum Gravity. The Renormalization Group scale is identified with the Tolman temperature for an isolated gravitational system in thermal equilibrium with Hawking radiation. As a result, the point of infinite local temperature is shifted from the classical black-hole horizon to the origin and coincides with a timelike curvature singularity. Close to the origin, the spacetime is determined by the scale-dependence of the cosmological constant in the vicinity of the Reuter fixed point: the free components of the metric can be derived analytically and are characterized by a radial power law with exponent $\alpha = \sqrt{3}-1$.
Away from the fixed point, solutions for different masses are studied numerically and smoothly interpolate between the Schwarzschild exterior and the scale-invariant interior.
Whereas the exterior of objects with astrophysical mass is described well by vacuum General Relativity, deviations become significant at a Planck distance away from the classical horizon and could lead to observational signatures. 
We further highlight potential caveats in this intriguing result with regard to our choice of scale-identification and identify future avenues to better understand quantum black holes in relation to the key feature of scale-invariance.
}

\begin{document} 
\maketitle
\flushbottom

\section{Introduction} \label{sec: Introduction}

Black holes are an essential and astrophysically relevant feature of the theory of General Relativity (GR), formulated by Einstein in the early 20th century. To date, theoretical predictions of Einstein's theory are in remarkable agreement with experimental observations~\cite{Berti:2015itd, LIGOScientific:2016aoc, EventHorizonTelescope:2019dse, LIGOScientific:2021sio}. Nevertheless, experiments leave room for deviations from GR and black holes provide a unique probe of the strong-field regime of gravity. In this context, two aspects of the black hole solutions in Einstein gravity are particularly intriguing.

First, stationary black holes in GR contain a curvature singularity~\cite{Hawking:1970zqf}. The respective unphysical divergence of curvature invariants results in the incompleteness of particle geodesics.\footnote{In the case of spinning black holes in GR, causality breaks down already before approaching the singularity, i.e., when crossing the inner Cauchy horizon.} These are signals that the theory is pushed beyond its regime of validity. At very short distances of the order of a Planck length, $l_{pl}\approx 10^{-35}$m, i.e.~at extremely high curvature scales, a theory of quantum gravity (QG) is expected to take over and provide a consistent microscopic description of spacetime.

Second, the stationary black hole solutions of GR are characterized by their event horizon. The event horizon is the portion of spacetime which is disconnected from future asymptotic infinity\footnote{In the stationary cases at hand the local definitions of event horizons, e.g.~as marginally trapped surfaces, agree with the global definition.}, in other words, from which no signal can escape. Event horizons are directly connected to recent observations including the quasinormal-mode ringdown of observed gravitational-wave signals~\cite{LIGOScientific:2016aoc} and a potential detection of the black-hole shadow, i.e., a characteristic central brightness depression in the intensity of very-long baseline interferometry~\cite{EventHorizonTelescope:2019dse}. Nevertheless, there is room in current experimental data for exotic compact objects as mimickers of black holes, see for instance gravastars~\cite{Mazur:2001fv,Visser:2003ge}, various versions of wormholes~\cite{Bronnikov:1973fh,Ellis:1973eft}, fuzzballs and firewalls~\cite{Susskind:1993ws,Mathur:2005zp,Almheiri:2012rt}, or, more recently, 2:2 holes~\cite{Holdom:2002xy,Holdom:2016nek,Holdom:2019bdv,Holdom:2019ouz,Ren:2019afg}. Potential observable signatures of horizon-like structures with a non-zero reflectivity include gravitational wave echoes associated with time-delayed signals following the main merger ringdown~\cite{Cardoso:2016rao,Abedi:2016hgu,Barcelo:2017lnx,Cardoso:2017njb,Oshita:2019sat} and excess emission in the central brightness depression of a black-hole shadow~\cite{Vincent:2020dij,Eichhorn:2022fcl,Carballo-Rubio:2022aed}.

For astrophysical objects of high mass, the curvature at classical horizon scales is many orders of magnitude below the Planck scale. Therefore a common belief is that QG does not play a role at horizon scales and drastic modifications of gravity at the classical level are needed to obtain spacetimes which substantially deviate from the GR predictions at horizon scales. Such a viewpoint may be presumptuous, given that the question of what are the fundamental degrees of freedom and which of these are relevant at different energies is an open question in QG, see~\cite{Eichhorn:2018yfc,Ashtekar:2021kfp,Surya:2019ndm,Brennan:2017rbf} for reviews on prominent candidate theories.
Different approaches to QG may lead to distinct expectations about the presence of horizon-scale modifications. Hence, observational insights into the existence or absence of horizons provides one of the rare opportunities of QG phenomenology. More generally, investigating phenomenological consequences of different approaches to QG constitutes an important step towards understanding the physical significance of the building blocks of the theory.

In the following, we adopt the viewpoint of asymptotic safety~\cite{Weinberg:1976xy,Weinberg:1980gg,Reuter:1996cp} (cf.~\cite{Bonanno:2020bil} for a recent review) as providing a high-energy (ultraviolet) completion for gravity and matter based on quantum scale invariance. Formally, quantum scale invariance corresponds to approaching a fixed-point of the renormalization group (RG) flow in the theory space of couplings. At the RG fixed point, dimensionless couplings become scale-independent. By dimensionless couplings, we refer also to the dimensionless counterpart of dimensionful couplings, obtained by multiplying with an appropriate power of the RG-scale. As a result, dimensionful couplings, such as the Newton coupling, acquire a characteristic high-energy scale-dependence.

We investigate phenomenological implications of such a scale-dependence of couplings, by means of a mechanism known as {\it RG-improvement}. Essentially, RG-improvement consists in retaining the dependence of some of the couplings on the RG scale and identifying the latter with a characteristic energy scale of the physical system in consideration. 
RG-improvement has been developed in the context of non-gravitational quantum-field theories~\cite{Coleman1973:rcssb} where it corresponds to a resummation of large logarithms and, for instance, can recover the quantum corrections to the Coulomb potential~\cite{Dittrich:2014eff}.
In the context of spacetime geometries, RG-improvement has been pioneered in~\cite{Bonanno:1998ye,Bonanno:2000ep} for the Schwarzschild geometry. Since then, RG improvement has been applied to black-hole physics~\cite{Bonanno:1998ye,Bonanno:2000ep,Bonanno:2006eu,Falls:2010he,Koch:2013owa,Kofinas:2015sna,Rincon:2018lyd, Rincon:2019cix} and in cosmology~\cite{Bonanno:2001xi,Bonanno:2001hi,Reuter:2005kb,Hindmarsh:2011hx,Bonanno:2018gck,Platania:2019qvo,Shapiro:2003ui,Sola:2013gha}, see also~\cite{Bonanno:2017pkg,Platania:2020lqb} for reviews. In RG-improvement of black-hole spacetimes, the RG-scale is typically identified with an appropriate notion of the local curvature scale and, as a result, could resolve classical curvature singularities~\cite{Bonanno:1998ye,Bonanno:2000ep,Torres:2014gta,Kofinas:2015sna,Torres:2017ygl,Adeifeoba:2018ydh}.
Potential phenomenological effects~\cite{Liu:2012ee, Eichhorn:2021iwq, Eichhorn:2021etc} face the challenge of overcoming the large separation between the Planck scale and the intrinsic scale of astrophysical black holes, see e.g.~\cite{Held:2019xde,Zhou:2020eth}. At the same time, the physical significance of the resulting regular black holes is currently challenged by a potential instability of the resulting inner Cauchy horizon~\cite{Carballo-Rubio:2018pmi,Bonanno:2020fgp}. In addition, there are indications that regular black holes which admit an asymptotic series expansion could be incompatible with a fundamental action principle~\cite{Knorr:2022kqp}.

We emphasize that QG phenomenology based on an RG-improvement of couplings does \emph{not} provide a unique prediction or strict first-principle derivation from asymptotic safety. First steps towards a first-principle derivation involve the physical RG running of couplings in terms of momentum-dependent correlation functions~\cite{Bosma:2019aiu}. The RG-improvement procedure is impacted by several physical assumptions.

First, RG-improvement depends inevitably on the specified scale-dependence of couplings. 
As of now, only the scale dependence of the Newton coupling and the cosmological constant has been investigated: The fixed-point scaling of the Newton coupling leads to resolution of the classical curvature singularity~\cite{Bonanno:2000ep}. However, the fixed-point scaling of the cosmological constant is expected to reintroduce a curvature singularity~\cite{Koch:2013owa, Pawlowski:2018swz, Adeifeoba:2018ydh}.

Second, the scale-dependence of couplings can be retained 
(i) at the level of the action~\cite{Reuter:2003ca,Reuter:2004nv}, 
(ii) at the level of the equations of motion~\cite{Bonanno:2001hi,Bonanno:2002zb,Babic:2004ev},
(iii) at the level of the metric of a classical solution~\cite{Bonanno:2000ep,Torres:2014gta,Kofinas:2015sna}, 
and, as more recently proposed, (iv) at the level of curvature invariants of a classical solution~\cite{Held:2021vwd}. In the cases (ii) and (iii), the RG-improvement is implemented in coordinate-dependent quantities and is thus additionally impacted by an unphysical dependence on the choice of coordinates~\cite{Held:2021vwd}.

Third, different scale identifications have been investigated. This includes a scale identification with curvature invariants, cf.~e.g.~\cite{Eichhorn:2021iwq, Held:2021vwd}, and a scale identification with geodesic distance to an asymptotic observer, cf.~e.g.~\cite{Bonanno:2000ep}.
Both essentially amount to an identification of the RG-scale with local curvature scales. As a result, classical GR remains a good approximation in regions of sub-Planckian local curvature. The present work highlights that this type of scale-identification is a physical assumption. Different physical assumptions can lead to qualitatively different results.
Insight into consistent choices of scale-identification might be gained from the Bianchi identities under the restrictive assumption of a separately conserved stress-energy tensor~\cite{Reuter:2003ca,Babic:2004ev,Domazet:2010bk}. The decoupling mechanism may provide another way of identifying the RG scale~\cite{Reuter:2003ca}.
\\

In the following, we implement the RG-improvement at the level of the action, and retain the scale-dependence of both the Newton coupling an the cosmological constant. Focusing on spherical symmetry, we investigate a scale-identification with local temperature instead of local curvature. The latter choice leads to qualitative differences in the results with regard to the existence of a horizon and therefore in particular with respect to the scale at which "QG effects" start to play a role.

Our choice of scale-identification is motivated by the field theory of a gravitating object surrounded by radiation in thermal equilibrium. The requirement of thermal equilibrium restricts the space of solutions to static spacetimes. For such spacetimes the RG-scale is identified with the local temperature of a stationary observer. The local temperature, introduced originally by Tolman and Ehrenfest~\cite{Tolman:1930etgr,Tolman:1930tes}, includes a blueshift-factor which diverges at the horizon of the classical black hole solutions. 

The obtained spacetimes describe horizonless geometries which match the classical Schwarzschild spacetime extremely well down to a Planck distance away from the classical horizon. Therefore, if they provide a description of the spacetime realized in nature up to a finite energy scale, in particular up to or beyond classical horizon scales, they may be expected to mimick classical black holes in all observables which are not sensitive to a Planck distance away from the horizon. Nevertheless, and as discussed above, the absence of a classical horizon structure may leave characteristic observational imprints on gravitational-wave signals or black-hole shadows~(e.g., see ~\cite{Cardoso:2019rvt} for a recent review).

The paper is organized as follows. In Sec.~\ref{subsec: RG-improvement idea}, we briefly review asymptotic safety as a motivation for including running couplings in the action. In Sec.~\ref{subsec: field equations}, we derive the general form of the modified field equations, whereas Sec.~\ref{sec: cutoff identification} is devoted to our choice of scale-identification. In Sec.~\ref{subsec: fixed-point analysis} and Sec.~\ref{subsec: numerical analysis}, spherically-symmetric solutions are studied analytically in the fixed-point regime and numerically at large distances, respectively. We finish with a discussion in Sec.~\ref{sec: discussion}.

\section{The modified Einstein-Hilbert theory}
\label{sec: Modified EH action}
\subsection{Running couplings in the classical action}
\label{subsec: RG-improvement idea}

This section reviews the origin and motivation behind an RG-improvement of couplings at the level of the classical action. Therefore, we introduce some of the ingredients of the asymptotic safety scenario~\cite{Weinberg:1976xy,Weinberg:1980gg,Reuter:1996cp} and the Wilsonian RG~\cite{Wilson:1971bg,Wilson:1971dh}.

The central object in the functional RG is the scale-dependent effective average action $\Gamma_k$~\cite{Wetterich:2001kra}, constructed formally from all operators which are compatible with the symmetries of the theory. In a gravitational context, these symmetries include diffeomorphism invariance and additional symmetries imposed on the matter sector. Starting from the bare classical action in the ultraviolet (UV), the effective average action at the scale $k$ is obtained by a momentum-shell integration of quantum fluctuations with momenta larger than the infrared (IR) cutoff $k$. In the limit $k\to 0$ the standard effective action is recovered. The scale-dependence of $\Gamma_k$ is governed by an exact functional RG flow equation~\cite{Wetterich:1992yh,Reuter:1996cp},
\begin{equation}\label{eq: flow equation}
k \partial_k \Gamma_k= \frac{1}{2} \text{STr}\qty[\qty(\Gamma_k ^{(2)}+\mathcal{R}_k)^{-1}k \partial_k \mathcal{R}_k],
\end{equation}
where $\Gamma_k ^{(2)}$ denotes the matrix of second functional derivatives of $\Gamma_k$ with respect to the fluctuation fields at fixed background. $\mathcal{R}_k$ is a regulator which suppresses IR modes, whereas its derivative in the flow equation results in the suppression for UV modes. $\text{STr}$ denotes a generalized functional trace. By construction, the main contribution to $\Gamma_k$ arises from momentum modes at the scale $k$. Although~\eqref{eq: flow equation} is an exact equation, finding solutions in the form of RG-trajectories requires approximation methods. For example, in a truncation of the theory space, the effective action is restricted to a finite sum of basis operators constructed from the fields and their derivatives. Local truncations of the effective action typically provide a good description for large values of $k$, whereas at low-momentum scales nonlocal terms become important and are difficult to handle~\cite{Capper:1973pv,Donoghue:1993eb}.

Here, the idea of RG-improvement comes into play. On the basis of the decoupling mechanism, a qualitative shortcut was proposed for the way from the UV to the IR~\cite{Reuter:2003ca}. The shortcut consists in including a spacetime-dependent cutoff $k=k(x)$ in the classical action, which has to be built from the properties of the classical physical system under consideration and therefore is expected to qualitatively capture the effect of some of the higher-order or nonlocal terms in the effective average action. The principle at work is that, at a certain decoupling scale, the physical parameters in $\Gamma_k$ may override the effect of the mathematical regulator appearing in the denominator of the trace in~\eqref{eq: flow equation}. As a consequence, $\Gamma_k$ at a finite scale below the cutoff scale does not deviate much from the standard effective action obtained in the limit $k\to 0$~\cite{Reuter:2003ca, Platania:2020lqb}. Identifying the decoupling scale might predict certain terms contained in the full effective action, but not in the original truncation. An example is given by the $\phi^4 \ln(\phi)$ term in the Coleman–Weinberg effective potential of massless $\phi^4$ theory~\cite{Coleman1973:rcssb}.
In light of the original motivation for RG-improvement inspired by asymptotic safety, the choice of cutoff should preserve the symmetries of the effective average action. In particular, imposing diffeomorphism invariance on the cutoff in an application of RG-improvement to gravity guarantees staying as close as possible to the theoretical framework of asymptotic safety. Nevertheless, the RG-improvement procedure does not provide a first-principle derivation from asymptotic safety even if such a symmetry requirement on $k$ is imposed. On the other hand, including characteristic information about the physical system is the key prerequisite for the interpretation and understanding of the phenomenological implications of the results obtained from RG-improvement. 

Although the original motivation for RG-improvement suggests that the scale-dependence of couplings should be implemented at the level of the action, RG-improvements at the level of the classical field equations or classical solutions have been employed widely (see~\cite{Koch:2014cqa, Platania:2020lqb} for reviews). Diffeomorphism-invariant scale-identifications with local curvature invariants or measures of proper distance, but also single component entries of the stress-energy tensor, as the energy density, have been used to identify the RG-scale. More recently, it has been proposed to perform the RG-improvement at the level of curvature invariants in order to guarantee a manifestly coordinate-invariant procedure~\cite{Held:2021vwd}.

Including leading-order quantum effects via a coordinate-dependent cutoff at the level of the action was explored for example in~\cite{Babic:2004ev, Domazet:2010bk, Koch:2010nn} for cosmology and in~\cite{Contreras:2013hua,Koch:2015nva} in the context of black holes. A more general proposal is to promote the cutoff to a function of the yet unknown metric, which generally leads to modified theories of gravity with structurally different equations than those of classical Einstein gravity.
\\

In what follows, scale-dependent couplings are included at the level of the classical action with an RG scale-identification motivated by the field theory of a gravity-matter system in thermal equilibrium. The following two subsections are devoted to the derivation of the modified field equations for general and for a concrete scale-dependence of the gravitational couplings in the action.

\subsection{General form of the field equations}\label{subsec: field equations}

Our starting point is the Einstein-Hilbert action
\begin{equation}
\label{eq: modified EH action}
S = \int \dd[4]{x}\sqrt{-g}\frac{1}{16 \pi G(k)}\qty(R - 2\Lambda(k)),
\end{equation}
where the Newton coupling and cosmological constant are considered as functions of the IR cutoff scale $k$. If $k=k(x)$ is a a scalar function on spacetime, not depending on the the metric, the equations obtained from a variation of the action with respect to the metric are given by~\cite{Reuter:2003ca} 
\begin{equation}
\label{eq: modified equations k metric-independent}
G_{\mu\nu} + \Lambda(k)g_{\mu\nu} = 8\pi G(k) T_{\mu\nu} + \Delta t_{\mu\nu},
\end{equation}
where $G_{\mu\nu}= R_{\mu\nu}-1/2\,R\,g_{\mu\nu}$ is the Einstein tensor and $T_{\mu\nu}$ is the energy-momentum tensor of the matter. The coordinate dependence of the Newton coupling introduces an additional effective energy-momentum tensor 
\begin{equation}
\Delta t_{\mu\nu} = G(k) \Big(\nabla_\mu \nabla_\nu - g_{\mu\nu}\Box\Big){G(k)}^{-1}.
\end{equation}
In situations where the energy-momentum tensor $T_{\mu\nu}$ is separately conserved, the contracted Bianchi identities and the scalar nature of $k$ impose a self-consistency condition on the cutoff function~\cite{Reuter:2003ca,Babic:2004ev,Domazet:2010bk}. Allowed cutoffs and solutions to~\eqref{eq: modified equations k metric-dependent} have been investigated in cosmological settings in~\cite{Babic:2004ev, Domazet:2010bk, Koch:2010nn} and in the context of black holes, see~\cite{Contreras:2013hua,Koch:2015nva}. It should be noted that in modified theories of gravity the conservation of the matter energy-momentum tensor at the classical and quantum level is not guaranteed. In our case no matter in the form of an energy-momentum tensor is included in the gravitational action.
\\

Let us now come to a key difference in our setup compared to previous studies of RG-improvement at the level of the action.
In the reviewed derivation of the equations~\eqref{eq: modified equations k metric-independent}, the RG scale $k$ is regarded as independent of the metric. In contrast, when varying the action, we retain the explicit variation of the RG scale $k$ with respect to the metric.
For a general metric-dependent cutoff function $k=k(g_{\mu\nu})$ the equations for the metric field are modified in comparison to~\eqref{eq: modified equations k metric-independent} by an additional term depending on the variation of the cutoff function with respect to the metric,
\begin{equation}\label{eq: modified equations k metric-dependent}
G_{\mu\nu} + \Lambda(k)g_{\mu\nu} = \Delta t_{\mu\nu} + \frac{\delta k}{\delta g^{\mu\nu}}\qty(\frac{1}{G(k)}\partial_{k}G(k)\qty(R-2\Lambda(k))+2\partial _k \Lambda(k)).
\end{equation}
To proceed with the study of solutions to this equation, we need to specify the RG-scale dependence of $G(k)$ and $\Lambda(k)$ in Sec.~\ref{subsec: running couplings} and the scale-identification $k=k(g_{\mu\nu})$ in Sec.~\ref{sec: cutoff identification}.

\subsection{Concrete scale-dependence of the gravitational couplings}\label{subsec: running couplings}

In this section we specify the scale-dependence of the gravitational couplings from asymptotic safety. The scale-dependence of the dimensionless Newton coupling $g(k) = G(k)k^2$ and of the cosmological constant $\lambda(k)= \Lambda(k)k^{-2}$ are dictated by their beta functions which are defined as the scale-derivatives $\beta_{g} = k \partial_k g(k) $ and $\beta_{\lambda} = k \partial_k \lambda(k)$. These can be computed in a given truncation of the effective average action, using functional RG methods. At a fixed-point, corresponding to a zero of the set of beta functions, the dimensionless gravitational couplings become constants $g_*$ and $\lambda_*$. The fixed-point is called interacting or non-Gaussian if at least one of its components is non-zero. According to asymptotic safety, there exists a non-Gaussian fixed-point of the RG-flow in the UV where the theory becomes quantum scale-invariant. Provided that the fixed-point comes with a finite number of relevant directions, the theory is rendered UV-complete and predictive. Evidence for the so-called Reuter fixed-point~\cite{Reuter:1996cp,Souma:1999at} is accumulating, cf.~for example~\cite{Bonanno:2020bil} for references and critical reflections. Throughout this paper, we shall assume that $g_*$ and $\lambda_*$ are positive. 

We specify to an explicit RG-scale dependence which connects the UV-scaling regime in the vicinity of the Reuter fixed point to the IR-scaling regime approaching (towards smaller $k$) the free Gaussian fixed point, cf.~\cite{Bonanno:2007wg, Gubitosi:2018gsl}. The underlying $\beta$-functions are derived in~\cite{Machado:2007ea}, see also App.~B in~\cite{Gubitosi:2018gsl}. An analytic expression which captures the two relevant scaling regimes can be derived by expanding the respective $\beta$-functions to second order in $g_k$. In terms of the dimensionful scale-dependent couplings $G(k)$ and $\Lambda(k)$, the resulting RG-scale dependence\footnote{Concretely, we refer to Eqs.~(4.3) and~(4.5) in~\cite{Gubitosi:2018gsl} with $k_0=0$ and $\Lambda(k=k_0)=0$. We set the effective dimensionful cosmological constant to zero at spatial infinity, as our analysis in Sec.~\ref{sec: spherically symmetric solutions} is focused on asymptotically flat spacetimes. While this is not exactly correct, the small curvature induced by Hawking radiation of a macroscopic black hole is quite small, e.g., $10^{-40}$ times cosmic mean density for a solar mass black hole.} reads
\begin{align}
    G(k) &= \frac{G_0}{\chi+(G_0/g_*)\;k^2}\;,
    \notag\\[0.5em]
    \Lambda(k) &= \frac{G_0\; (\lambda_*/g_*)\;k^4}{\chi + (G_0/g_*)\;k^2}\;.
    \label{eq: scale dependence}
\end{align}
Here, we have introduced the fixed-point values $g_*$ and $\lambda_*$ and a fiducial dimensionless parameter $\chi$. The explicit values for $g_*$ and $\lambda_*$ which match \cite{Gubitosi:2018gsl} can be obtained by taking the respective limits of Eqs.~(4.3) and~(4.5) in~\cite{Gubitosi:2018gsl} and fixing $\chi=1$. For $\chi=0$, one recovers the UV scaling regime, i.e.,
\begin{align}
    G(k) \sim k^{-2}\;,
    \quad\quad\quad
    \Lambda(k) \sim k^2.
\end{align}
For small $k^2\ll\chi$ (and $\chi=1$), the couplings scale towards the free IR fixed point, i.e.,
\begin{align}\label{eq: IR limit}
    G(k) \rightarrow G_0\;,
    \quad\quad\quad
    \Lambda(k) \sim k^4.
\end{align}
General Relativity with vanishing cosmological constant is recovered in the limit $\chi= 1$ and $\omega \equiv 1/g_* = 0$.
We introduce $\chi$, such that small $\chi>0$ allows us to perturb the fixed-point equations. We will make use of an expansion of the field equations around $\chi=0$ in our analysis of solutions to the field equations in the fixed-point regime, cf.~Sec.~\ref{subsec: fixed-point analysis}. In the following, in addition to $\hbar = c= 1$, the classical Newton constant is set to $G_0 = 1$, which implies that dimensionful quantities are measured in Planck units.

The $k$-dependence~\eqref{eq: scale dependence} illustrates the weakening of the dimensionful Newton coupling at high energies, whereby the transition scale to the quantum gravity regime is typically associated with the Planck scale. It was shown some time ago that such an anti-screening behaviour of the Newton coupling might lead to the resolution of classical singularities in the context of an RG-improvement~\cite{Bonanno:2000ep}. 

In an RG-improvement of the classical Schwarzschild (Anti-) de Sitter solutions, however, the quadratic divergence of the cosmological constant in the UV reintroduces the curvature singularity~\cite{Koch:2013owa, Pawlowski:2018swz} at the center. A more detailed analysis of conditions for black hole singularity resolution was performed in \cite{Adeifeoba:2018ydh}. As a result, a finite cosmological constant in the IR might be compatible with singularity resolution, provided that it vanishes fast enough in the UV.

\section{Scale-identification from thermal field theory}\label{sec: cutoff identification}

The IR cutoff $k$ in the definition of the effective average action $\Gamma_k$ is a mathematical parameter associated with the RG and the flow equations. Integrating out all the quantum fluctuations from the effective average action, the $k$-dependence cancels out, such that the full physics can be extracted from the field equations governed by the effective action $\Gamma_0$. In the context of RG-improvement, however, the integration is extended only down to finite $k$. Therefore, in physical applications of RG-improvement -- especially in the search of solutions to the field equations~\eqref{eq: modified equations k metric-dependent} -- the cutoff parameter must be related to a characteristic energy scale of the system. This procedure is referred to as scale-identification. 
In highly symmetric frameworks, such as homogeneous and isotropic cosmology or spherically-symmetric black-hole spacetimes, the scale is typically identified with local curvature scales of the classical solutions.
In~\cite{Platania:2019kyx} an iterative RG-improvement was suggested based on a self-adjusting cutoff, which takes into account the backreaction effects due to corrections to the Einstein equations arising from a running Newton coupling. 
More generally, RG-scale settings at the level of the action and equations were proposed in~\cite{Reuter:2003ca,Babic:2004ev,Domazet:2010bk,Koch:2010nn,Domazet:2012tw,Koch:2014joa} based on diffeomorphism invariance and the Bianchi identities. Alternatively, the decoupling mechanism in effective field theories may serve as a route towards a self-consistent scale-identification~\cite{Reuter:2003ca}.
\\

In what follows we introduce a metric-dependent cutoff function motivated by the Euclidean field theory of a finite-size gravity-matter system in thermal equilibrium. According to Hawking~\cite{Hawking:1975vcx}, a semi-classical black hole emits black-body radiation at a finite temperature. Putting the system consisting of the black hole and the radiation in an isolated box with finite volume, the requirement of thermal equilibrium implies that physical properties of the system cannot depend on time from the viewpoint of observers at rest with respect to the system. As a consequence, the spacetime must be stationary. Restricting to a system with a finite spatial volume $V$ in a static spacetime, the metric can be written as
\begin{equation}
\dd{s^2} = g_{00}(x^i)\dd{t^2} + h_{ij}\dd{x^i}\dd{x^j},
\end{equation}
where $g_{00}<0$ and $h_{ij}$ do not depend on the coordinate $t$. The latter is proportional to the proper time of stationary observers and can be fixed by identification with the proper time of a particular observer. Then the gravitational redshift factor between that observer and another one at $x^\mu$ is given by $\sqrt{-g_{00}(x^i)}$. The radiation is taken into account by considering a relativistic field with appropriate boundary conditions imposed at the walls of the box. In thermal equilibrium, the product of the local temperature of an infinitesimally small subsystem and the redshift factor of the gravitational field remains constant. This fact, also known as the Tolman-Ehrenfest effect~\cite{Tolman:1930tes}, expresses the dependence of the proper temperature of a local observer on the gravitational potential at the point where the measurement is made. In turn, if $T$ denotes the temperature of the black-body radiation measured by a stationary observer at infinity in the limit where $V\to \infty$, then $T/ \sqrt{-g_{00}(x^i)}$ defines the local temperature for an observer at the point $x^\mu$. On this basis a cutoff function is constructed as follows,
\begin{equation}\label{eq: cutoff identification}
k = \frac{T}{\sqrt{-g_{00}(x^i)}}.
\end{equation}
In other words, the momentum cutoff at a certain point in the confined spatial volume is identified with the local (blueshifted) value of the physical observable whose expectation value is $T$ for a stationary observer at infinity. For classical black holes, $T$ is associated with the the characteristic temperature of the emitted black body spectrum. We should note that the RG-improvement is started by promoting the couplings in the classical action to running couplings depending on the RG-scale $k$. Therefore, the corresponding classical solutions play a distinguished role in our setup and solutions to the RG-improved theory should be understood as potential quantum modifications to their classical counterparts. In particular, the identification of $k$ with the temperature of the classical spacetime, combined with the running $\propto k^4$ of the cosmological constant~\eqref{eq: IR limit} implies that the quantum-modified spacetime automatically encodes the effect of radiation with temperature $T=T_{\text{Hawking}}$. In addition, as $k$ is defined to be proportional to the temperature, it scales with inverse powers of the
mass of the classical object. Therefore, for macroscopic black holes, the RG-scale parameter $k$ at spatial infinity will be very small. From its finite value, however, we expect an insensitivity of the quantum-modified solution to IR effects.

For a classical black hole, the redshift factor goes to zero at the event horizon, where $g_{00} = 0$, and therefore the local temperature diverges. The cutoff as defined in~\eqref{eq: cutoff identification} therefore diverges at a finite distance away from the classical singularity. Let us note that previous studies of RG-improvement in spherical symmetry typically identify the RG cutoff-scale with local \emph{curvature} or \emph{proper distance} scales of the classical theory. Nevertheless, from the viewpoint of the thermal field theory of a static gravitational field interacting with its environment in an isolated region of space, it is the local \emph{temperature} which sets the physical energy scale.

\begin{figure}[t]
\centering
\includegraphics[width=.3\textwidth]{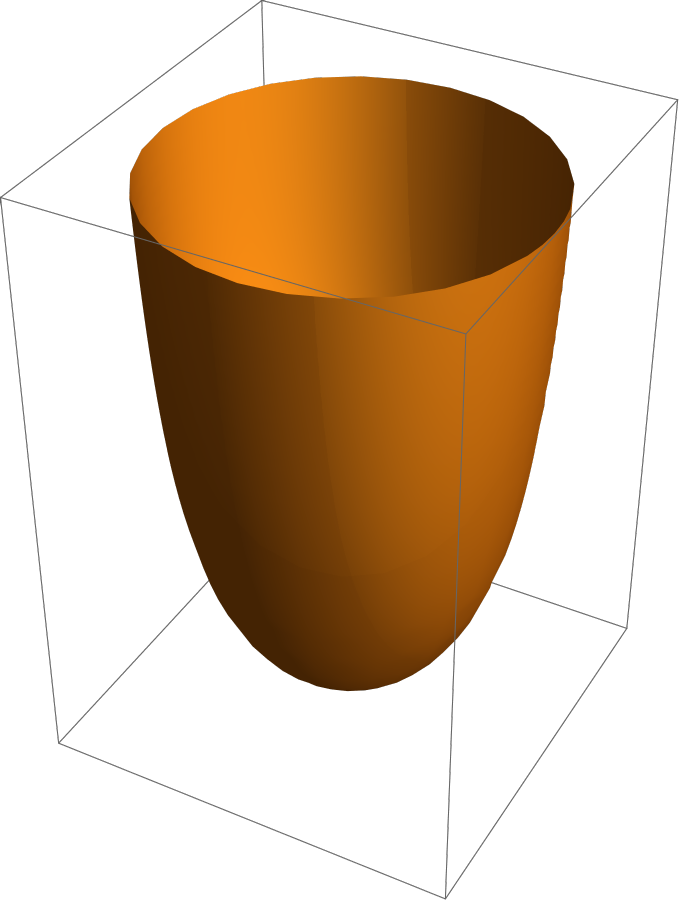}
\hspace{1cm}
\includegraphics[width=.3\textwidth]{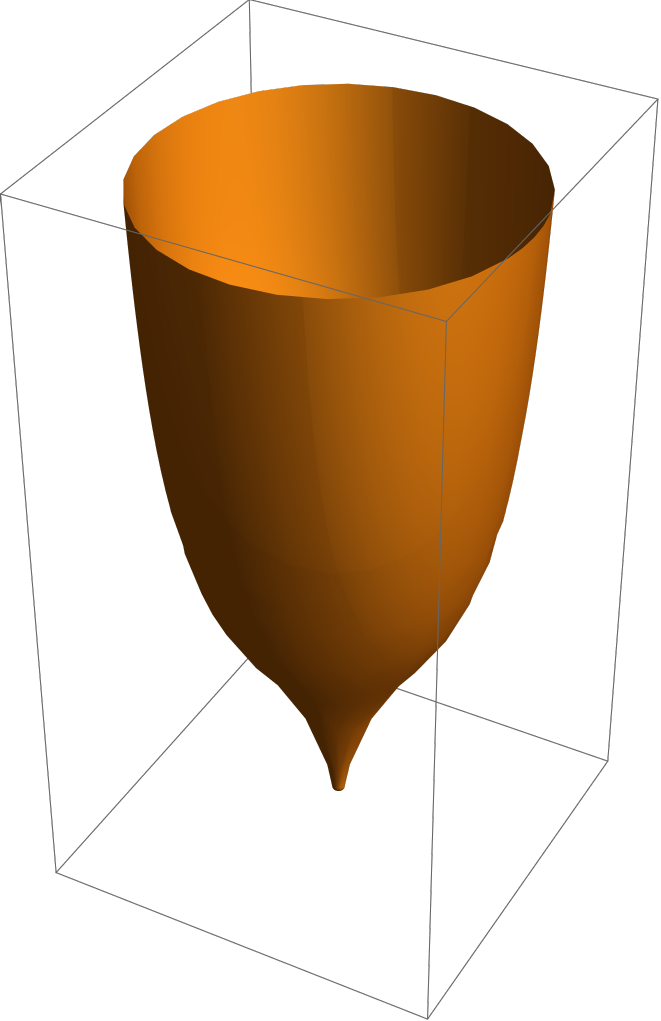}
\caption{\label{fig:cigar} The geometry of the $(t,r)$ Euclidean black hole manifold, a.k.a. ``cigar'' in General Relativity (left) and its RG-improvement in this study (right), for $M\simeq 1$ (in Planck units).   }
\end{figure}

To illustrate the origin of the parameter $T$ for classical black holes and its role in the definition of the cutoff, it is useful to revisit the central elements in the derivation of the Hawking temperature in the framework of Euclidean path integrals. Originally, Hawking's derivation of black hole radiance~\cite{Hawking:1975vcx} utilized methods from quantum field theory on curved background. The result, that a black hole radiates at finite temperature, suggests an equivalent treatment in terms of Euclidean field theory and thermal Green's functions~\cite{Gibbons:1976ue, Gibbons:1976pt, Hawking:1978jz}. 
Let us consider a static spherically symmetric spacetime of the form
\begin{equation}\label{eq: metric with f(r)}
\dd{s^2} = -f(r)\dd{t^2} + g(r)^{-1}\dd{r^2} + r^2\dd{\Omega^2},
\end{equation} 
where $\dd{\Omega^2}$ is the infinitesimal area element on the $S^2$. We assume that the spacetime has an event horizon at $r_h$ where $f(r_h)=g(r_h)=0$ and where the derivates of the metric functions do not vanish. A well-known example is the Schwarzschild spacetime with lapse function $f(r) = g(r) =1-2M/r$ and $r_h=2M$ which describes the exterior of a spherically symmetric object of radius $r_h$ and mass $M$ measured at infinity. The coordinate singularity at $r=r_h$ can be removed by a coordinate transformation to Eddington-Finkelstein coordinates. In the context of Euclidean path integrals for thermal systems, the positive definite Euclidean metric is defined by a complexification of the time coordinate, $t\to i\tau$. The canonical partition function for the gravitational field is consequently written as a sum over all smooth Riemannian geometries which are periodic in imaginary time, $\tau\to \tau + \beta$, with period $\beta = T^{-1}$ at infinity,
\begin{equation}
Z(\beta) = \int \dd{[g_{\mu\nu}]}e^{-S_E \qty[g_{\mu\nu}]},
\end{equation}
where $S_E$ is the Euclidean action. In particular, our scale identification~\eqref{eq: cutoff identification} retains the global diffeomorphism-invariance of the Euclidean action within the class of thermal Euclidean manifolds, i.e.~those with a timelike Killing vector and periodicity of $\tau \to \tau+\beta$. Within this class, $l= 1/k = \beta \sqrt{g_{00}(r)}$ has a covariant interpretation as the geometric circumference of the Euclidean manifold at radius $r$.
Taylor-expanding the metric functions in the near-horizon limit, i.e.~for small $(r-r_h)$, to first order the metric looks locally like Rindler space. After a coordinate transformation $(\tau, r)\to (\phi, \rho)$, where $\phi = \sqrt{\abs{f'(r_h)g'(r_h)}/4}\tau$ and $\rho^2 = 4 (r-r_h)/g'(r_h)$, the metric in a neighborhood of the horizon can be written as
\begin{equation}\label{eq: metric local horizon limit}
\dd{s^2} = \dd{\rho^2} + \rho^2\dd{\phi^2},
\end{equation}
where we have omitted the two extra dimensions corresponding to the $S^2$ in~\eqref{eq: metric with f(r)}. Smoothness of the metric~\eqref{eq: metric local horizon limit} requires the identification of $\phi$ as an angular variable with period $2\pi$ and restriction of the range of $r$ to $r\geq r_h$. The last condition arises from the fact that the Killing vector $\partial_\tau$ vanishes at $r=r_h$. Then the line element becomes that of a flat disc with radius $\rho$ and polar angle $\phi$. This in turn fixes the imaginary time period $\beta$ and thereby the temperature $T$ measured by a stationary observer at infinity,
\begin{equation}\label{eq: T}
T = \sqrt{\frac{\abs{f'(r_h)g'(r_h)}}{16 \pi^2}}.   
\end{equation}
For a Schwarzschild black hole, $T = T_H = 1/8\pi M$ reproduces the Hawking temperature. The associated Euclidean manifold has topology $S^1 \cross \mathbb{R_{+}}$ and looks like the surface of a cigar with a smoothly closed end on the one side, while asymptotically approaching a cylinder at large $r$ (Figure \ref{fig:cigar}, left). With this picture in mind,~\eqref{eq: cutoff identification} identifies the characteristic length scale $l= 1/k$ with the radius of the cigar. In the UV limit $k\to \infty$, the point $r=r_h$ is reached. It is a fixed-point of the isometry $\tau \to \tau + \beta$ and represents the Euclidean continuation of the bifurcate Killing horizon in the Lorentzian black-hole solution~\cite{Wald:1995yp,Ross:2005sc}. Since $k \to \infty$ at this point, the RG-improvement using \eqref{eq: cutoff identification} can change the geometry significantly (Figure \ref{fig:cigar}, right). 

In our situation, inferring the physical meaning of the parameter $T$ for the quantum black hole is not straightforward. The dependence of the cutoff function~\eqref{eq: cutoff identification} on the $(0,0)$-component of the metric results in modified field equations whose solutions will in general differ from the classical GR solutions. In particular, it is a priori not clear whether these spacetimes exhibit a horizon. In a complete treatment one would aim to solve a variational problem for the metric components such that the Euclidean effective action is minimized. Nevertheless, in the context of RG-improvement introduced in Sec.~\ref{subsec: RG-improvement idea}, we expect the shortcut of introducing a local-temperature-dependent cutoff to capture qualitative features of nonlocal terms and radiative corrections contained in the full effective action for the thermal system. On these grounds, there are different ways of dealing with the parameter $T$. If the new spacetime is found to contain an event horizon with the necessary conditions for the derivation of~\eqref{eq: T}, then it would be consistent to identify $T$ with the temperature~\eqref{eq: T}. However, even in spherical symmetry, the full modified field equations can only be studied numerically. Therefore, in practice, as in other forms of RG-improvement, a classical input must be made at some stage of the procedure. The running couplings in~\eqref{eq: scale dependence} are defined such, that they approach $G=G_0\equiv1$ and $\Lambda = 0$ in the IR. Consequently, in the spherically symmetric case, the Schwarzschild solution at large distances can be used as a classical input. Then $T = T_H = 1/8\pi M$ provides a consistent identification for the numerical solution at large distances. Whereas we employ this identification in the numerical analysis of solutions in Sec.~\ref{subsec: numerical analysis}, we shall in general consider $T$ as a free parameter.

\section{Spherically symmetric solutions in different regimes}\label{sec: spherically symmetric solutions}

In the following, we limit the search for solutions to the modified Einstein-Hilbert theory discussed in Sec.~\ref{sec: Modified EH action} and Sec.~\ref{sec: cutoff identification} to static spherically symmetric spacetimes, written in the form
\begin{equation}\label{eq: metric ansatz}
\dd{s^2} = -A(r)\dd{t^2} + B(r)\dd{r^2} + r^2\dd{\Omega^2}.
\end{equation}
To derive the field equations governing the metric functions $A$ and $B$, we may follow two different routes: On the one hand, we can insert the ansatz~\eqref{eq: metric ansatz} into the general field equations~\eqref{eq: modified equations k metric-dependent}.\footnote{We note that the $(2,2)$ and $(3,3)$-components of the field equations~\eqref{eq: modified equations k metric-dependent} are redundant once $A(r)$ and $B(r)$ are governed by the $(0,0)$- and $(1,1)$-components of the field equations. Hence, there are only 2 nontrivial field equations, as for the derivation following~\eqref{eq: action variation}.} On the other hand, we can insert the ansatz~\eqref{eq: metric ansatz} into the modified Einstein-Hilbert action~\eqref{eq: modified EH action} and directly compute the variations
\begin{equation}
\label{eq: action variation}
\frac{1}{\sqrt{-g}}\frac{\delta S}{\delta A} = 0 \quad \text{and} \quad \frac{1}{\sqrt{-g}}\frac{\delta S}{\delta B} = 0.
\end{equation}
In both cases, we use the scale-dependent running couplings in~\eqref{eq: scale dependence} and the cutoff identification in~\eqref{eq: cutoff identification}. We confirm that both derivations result in the same field equations.

The cutoff function introduced in Sec.~\ref{sec: cutoff identification} depends only on the metric and not on its derivatives. Hence, the field equations remain of second order, as in GR. 

For the ansatz in~\eqref{eq: metric ansatz}, the radial field equation can be solved algebraically to give $B(r)$ in terms of $A(r)$ and its derivatives, cf.~App.~\ref{app: differential equations} for details. Therefore, solutions are completely characterized by a second-order nonlinear ordinary differential equation for $A(r)$,
\begin{equation}
\label{eq: eom}
A'' + f(r,A, A')=0.
\end{equation}
The explicit form of $f(r,A,A')$ is given in App.~\ref{app: differential equations}.

\subsection{Exact solutions to the fixed-point equations}\label{subsec: fixed-point analysis}

\begin{figure}[t]
\centering
\includegraphics[width=0.55\textwidth]{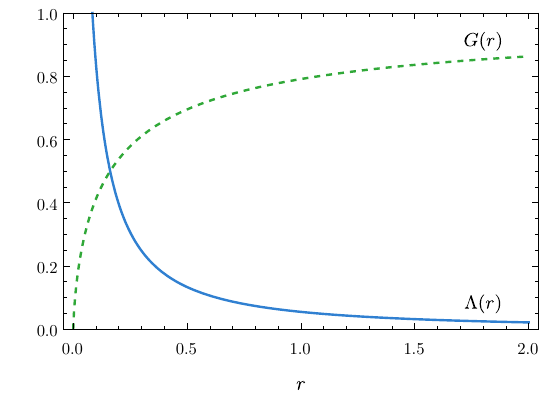}
\caption{\label{fig: running couplings as function of k} Dimensionful Newton coupling (green) and cosmological constant (blue) as functions of the radial coordinate, dictated by the scale-dependence of the couplings according to~\eqref{eq: scale dependence} ($\chi = 1$) and inserting the power law solution $A \propto r^{\sqrt{3}-1}$ to the field equations in the fixed-point regime. We set $\omega = 1$, $\lambda_* = 1$ and $T = 1/8\pi M$ with $M=1$.}
\end{figure}

In this section, we shall study the differential equation for the $(0,0)$-component of the metric in the fixed-point regime, i.e.,~with the gravitational couplings running according to~\eqref{eq: scale dependence} with $\chi=0$. Setting $\chi = 0$ in the differential equation for the metric function $A$, cf.~App.~\ref{app: differential equations}, the equation becomes independent of the UV fixed-point value of the dimensionless Newton coupling. Consequently, the running of the cosmological constant is the only relevant ingredient for solutions to the fixed-point theory at high energies. In addition, the equation depends only on the parameter combination $\lambda_* T^2$, which can be traced back to the $k$-dependence of the dimensionful couplings at the fixed-point and our definition of the cutoff-function.

The momentum-cutoff $k\propto A^{-1/2}$ becomes large when the metric function $A$ is close to zero. A zero at a finite radius would correspond to a horizon. In contrast, we find that the fixed-point equations admit power-law solutions around $r=0$,
\begin{equation}\label{eq: power laws}
A(r) \propto r^\alpha.
\end{equation}
For a special choice of the proportionality constant in~\eqref{eq: power laws}, a parabolic solution is given by $A(r) = 2 \lambda_* T^2 r^2$. A trivial solution to the equations in the fixed-point regime is given by~\eqref{eq: power laws} with the power $\alpha = 0$, i.e.,~a constant solution for which the radius of the counterpart to the Euclidean cigar never shrinks to zero and therefore never reaches the fixed-point. Additionally, there are solutions with powers $\alpha = \pm \sqrt{3}-1$. The latter solution with the negative sign diverges at $r=0$ and is not appropriate to describe the UV regime $k\to \infty$. Therefore, except for the fine-tuned quadratic solution there is a distinct solution $A=a_0 r^{\alpha}$ with power-law exponent
\begin{equation}
\alpha = \sqrt{3}-1.
\end{equation}
Our numerical analysis in Sec.~\ref{subsec: numerical analysis} shows that this is the correct power law for $A$ at small $r$.
Inserting the power-law solution for $A$, together with the fixed-point scaling of $G$ and $\Lambda$, into the expression for $B$ given by App.~\eqref{eq: B rule}, leads to $B=0$. On the other hand, the numerical solutions presented in Sec.~\ref{subsec: numerical analysis} show that in a sufficiently small neighbourhood of the origin $B$ is given by a positive power of the radial coordinate. The leading non-zero term for $B$, representing an approximate solution to the fixed-point theory, can be obtained by perturbing the fixed-point equations. Concretely, we consider small $\chi >0$ corrections to the exact solution, $A(r) = a_0 r^{\sqrt{3}-1} + \chi a_1 r^{\alpha_1} $, and expand~\eqref{eq: general differential equation for A} around $\chi=0$ to lowest non-trivial order.
As a result, $B$ receives corrections proportional to $\chi^2$ with a leading $r$-dependence 
\begin{equation}\label{eq: power laws B}
B(r) \propto r^{2\alpha}.
\end{equation}
The proportionality constant $b_1$ is fully fixed by $\chi^2$ times a factor depending on the proportionality constant $a_0$ in the expression for $A$, and on the free parameters $T$ and $\omega$, cf.~App~\ref{app: differential equations}.

\begin{figure}[t]
\centering 
\includegraphics[width=0.6\textwidth]{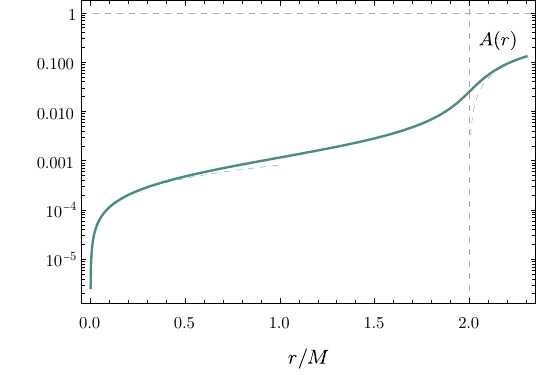}
\caption{\label{fig: A numerical analytical} Numerical solution to the field equations (green) according to the interpolating scale-dependence~\eqref{eq: scale dependence} of the gravitational couplings ($\chi=1$) . For the numerical integration, Schwarzschild initial conditions are imposed at $r_i = 10M$, we set $\lambda_*=1$, $\omega = 1$ and $T = 1/8\pi M$ with $M=1$. The asymptotic solution in the fixed-point regime is given by $A\propto r^{\alpha}$ and receives $r^{2 \alpha}$-corrections in a finite neighborhood of $r=0$ (blue dashed). For comparison, the classical Schwarzschild solution is shown for $r/M>2$ (gray dashed).}
\end{figure}

The first observation is that the renormalization of the couplings in the action with a thermal scale setting function, leads to a finite metric at the origin. Fig.~\ref{fig: running couplings as function of k} shows the running Newton coupling,~\eqref{eq: scale dependence} with $\chi=1$, as a function of the radial coordinate for the critical solution. It vanishes at the origin, reflecting the anti-screening property of gravity in the UV, whereas at large distances from the origin the classical Newton constant is recovered. On the other hand, the effective cosmological constant becomes negligible in the IR, but diverges at the center. 

Different from the classical Schwarzschild black hole, the fixed-point of infinite local temperature signalling the existence of a horizon in the thermal theory, is shifted towards the center. This result is non-trivial, in fact, a power law solution $A\propto (r-r_h)^\alpha$ at a horizon at finite $r=r_h$ could have been realized as well.
Put differently, we find that taking the Euclidean thermal Schwarzschild black bole as a guiding principle for the scale-identification, the outcome describes a spacetime where the fixed-point of the underlying isometry is now at the origin of the radial coordinate. In turn, the infinite temperature-blueshift is realized only in the limit of $r\to 0$, which coincides with the limit of approaching the UV fixed-point using measures such as the proper distance or local curvature of the classical spacetime. In the above sense, we are dealing with a spacetime for which various measures of energy such as the blueshift factor, the local curvature or inverse powers of the proper distance diverge at the same spatial point in the UV limit. 

In the presence of a running cosmological constant, a curvature singularity is expected at the origin~\cite{Adeifeoba:2018ydh}. In fact, various curvature scalars such as the Ricci scalar, the Weyl scalar and the Kretschmann scalar diverge. As an example, the leading term of the Kretschmann scalar at small $r$ is given by
\begin{equation}\label{eq: Kretschmann}
R_{\mu\nu\rho\sigma}R^{\mu\nu\rho\sigma}\propto r^{-4\sqrt{3}}.
\end{equation}
Compared to the scaling $R_{\mu\nu\rho\sigma}R^{\mu\nu\rho\sigma}|_\text{Schwarzschild}\propto r^{-6}$, the classical singularity is strengthened.
To investigate the causal structure of the spacetime globally, in the following Sec.~\ref{subsec: numerical analysis}, we study numerical solutions away from the fixed-point.

\subsection{Numerical solutions at large distances}\label{subsec: numerical analysis}

\begin{figure}[t]
\centering
\includegraphics[width=.49\textwidth]{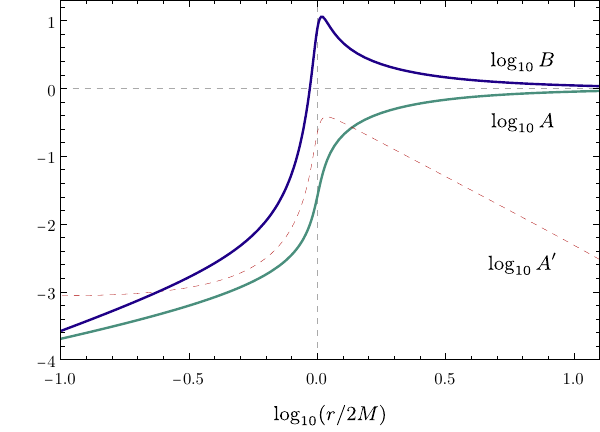}
\hfill
\includegraphics[width=.49\textwidth]{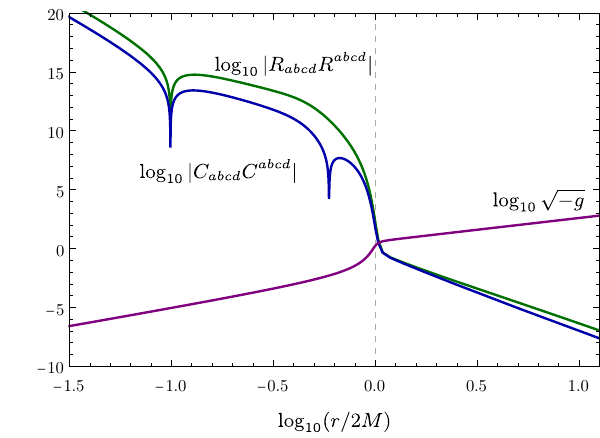}
\caption{\label{fig: numerical result} Numerical solution to the field equations obtained with scale-dependent gravitational couplings~\eqref{eq: scale dependence} ($\chi=1$) and scale-identification~\eqref{eq: cutoff identification}. Schwarzschild initial conditions are imposed at $r_i = 10M$, we set $\lambda_*=1$, $\omega = 1$ and $T = 1/8\pi M$ with $M=1$. Left: Metric function $A$ (green), $B$ (purple) and derivative $A'$ (red dashed). Right: Volume element (purple), Kretschmann scalar (green) and squared Weyl tensor (blue).}
\end{figure}

In the previous section, we derived analytical solutions to the field equations~\eqref{eq: modified equations k metric-dependent} in the fixed-point regime where $G$ and $\Lambda$ scale according to~\eqref{eq: scale dependence} with $\chi=0$. More generally, the scale-dependence of the couplings, interpolating between the UV and the IR, can be approximated via~\eqref{eq: scale dependence} with $\chi =1$. The effective cosmological constant at infinity is set to zero, allowing us to search for asymptotically flat solutions. To obtain numerical solutions away from the fixed-point, we transform to a system of two first-order differential equations and set $\chi=1$ in the field equations, cf.~App.~\ref{app: differential equations}. As was pointed out in Sec.~\ref{sec: cutoff identification}, a consistent identification and interpretation of the free parameter $T$ is not straightforward. For the numerical integration, $T$ is identified with the Hawking temperature of a Schwarzschild black hole, i.e.~$T= T_H = 1/8\pi M$. We make this choice, since the classical Schwarzschild spacetime is used to set initial conditions at large distances $r_{i}\gg r_h = 2M$. The parameters $\lambda_*$ and $\omega$, corresponding to the UV fixed-point value of the dimensionless cosmological constant and the inverse UV fixed-point value of the dimensionless Newton coupling, are of order unity in asymptotic safety. The result of the integration for an object of Planck mass is shown in Fig.~\ref{fig: numerical result}.

\begin{figure}[t]
\centering
\includegraphics[width=.49\textwidth]{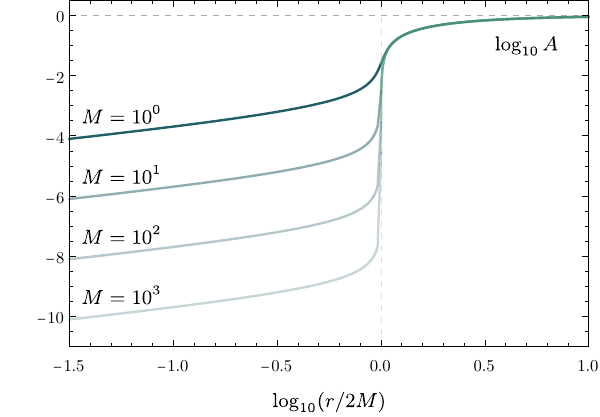}
\hfill
\includegraphics[width=.49\textwidth]{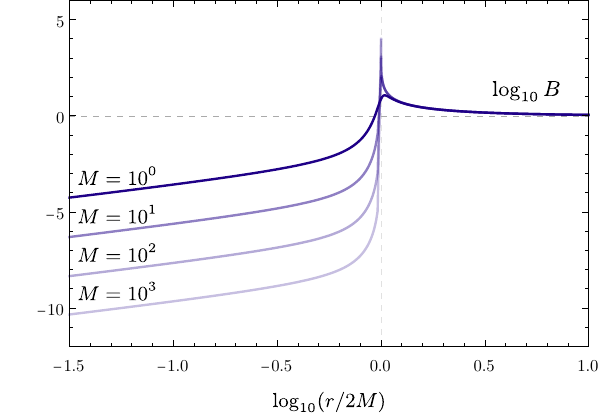}
\caption{\label{fig: different masses} Metric functions $A$ (left) and $B$ (right) of for different masses. At small $r$ the solutions are given by power laws $A\propto a_0(M)r^{\alpha}$ and $B\propto b_1(M)r^{2\alpha}$ with $\alpha = \sqrt{3}-1$, $a_0(M) \propto M^{-\sqrt{3}-1}$ and $b_1(M)\propto M^{-2\sqrt{3}}$.}
\end{figure}

The function $A$ associated with the $(0,0)$-component of the metric matches the Schwarz-schild solution well down to a Planck distance away from the classical horizon, which can also be seen from Fig.~\ref{fig: A numerical analytical}. In the interior, deviations from GR are significant. In the intermediate transition region, the radial metric function $B$ and the derivative $A'$ are peaked. The vanishing second derivative $A''$ can be regarded as a phase transition between the Schwarzschild geometry and the new spacetime obtained from the modified field equations. Our numerical results show power-law behaviour at small $r$, close to the origin. In particular, they allow us to single-out the correct exponent from the allowed ones presented in Sec.~\ref{subsec: fixed-point analysis}. This power-law exponent is $\alpha=\sqrt{3}-1$ for $A$, which gives rise to $\beta = 2 \alpha$ as the asymptotic solution to the power-law behavior of $B$. The mass-dependence of the two proportionality constants in the power-law expressions can be approximated numerically by investigating solutions with different masses, as shown in~Fig.~\ref{fig: different masses}. We find $a_0(M) \propto M^{-\sqrt{3}-1}$ for the proportionality constant in the power law for $A$, and $b_1(M)\propto M^{-2\sqrt{3}}$ for the proportionality constant in the power law for $B$. With increasing mass, the transition between the classical and the interior region is sharpened. However, the metric functions remain smooth, in contrast to the classical Schwarzschild spacetime for which our choice of coordinates is singular at the horizon.
For astrophysical objects with mass of the order $M_\odot \sim 10^{38}$, deviations from a classical horizon would be tiny. Nonetheless, they could give rise to observable gravitational wave signatures in the form of late-time echoes~\cite{Abedi:2016hgu,Barcelo:2017lnx,Cardoso:2017njb}. 
Different parameter values for $\lambda_*$ and $\omega$ impact the solutions as expected from~\eqref{eq: scale dependence}, i.e.~the limit $\omega = 0$ reproduces the Schwarzschild solution, whereas a finite $\omega > 0$ results in a horizonless spacetime. Deviations from the Schwarzschild geometry outside the classical horizon become larger as $\omega$ is increased. The result is similar if higher values for $\lambda_*$ are specified.

As anticipated from the analytical study in Sec.~\ref{subsec: fixed-point analysis}, the spacetimes exhibit a curvature singularity where the Kretschmann scalar scales as $\sim r^{-4\sqrt{3}}$ at the center. The curvature singularity is not hidden behind a classical event horizon as for the Schwarzschild black hole, but is instead naked and timelike. At this stage it is unclear whether such a singularity is physically relevant, keeping in mind that the nature of observables in quantum gravity in the UV is indeterminate. 

\section{Discussion and future prospects}
\label{sec: discussion}

Inspired by Weinberg's asymptotic safety scenario for quantum gravity~\cite{Weinberg:1976xy,Weinberg:1980gg}, we derive modified field equations for general RG scale-dependent gravitational couplings in the Einstein-Hilbert action. In contrast to previous work, we include variations of the RG-scale $k$ with respect to the metric. Adopting the viewpoint of Euclidean field theory of an isolated system consisting of a static black hole in thermal equilibrium with Hawking radiation~\cite{Gibbons:1976ue, Gibbons:1976pt, Hawking:1978jz}, we construct a thermal scale-identification which associates $k$ with the local temperature~\cite{Tolman:1930etgr, Tolman:1930tes} as measured by a stationary observer. 
We specify to an RG-scale dependence of the gravitational couplings which interpolates between the scaling regime of asymptotically safe gravity in the ultraviolet (UV), i.e., for large $k$, and the scaling regime approaching the free (Gaussian) fixed point in the infrared (IR), i.e., for small $k$. In the IR, the couplings scale like $G(k)\sim\text{const}$ and $\Lambda(k)\sim k^4$, implying that in this regime the field equations are equivalent to GR sourced by thermal radiation. Our approximation neglects a subsequent scaling-regime in the deep IR, in which the dimensionful cosmological constant freezes out to its observed value. Thereby, asymptotically flat spacetimes can be studied numerically. Assuming that at large radii the classical Schwarzschild spacetime is recovered, the free parameter in our scale-identification is identified with the Hawking temperature.
Our numerical solutions, obtained by imposing the Schwarzschild metric in the IR and integrating towards smaller distances, reproduce vacuum General Relativity in the exterior region remarkably well for large masses. Deviations of the quantum counterparts from classical black holes become significant only at a Planck distance away from the classical horizon. The resulting RG-improved spacetime would thus mimick classical black holes in observations which do not explicitly probe the horizon as a perfect absorber. On the one hand, such quantum modifications might lead to observational signatures, such as gravitational wave echos~\cite{Cardoso:2016rao,Abedi:2016hgu,Barcelo:2017lnx,Cardoso:2017njb,Oshita:2019sat} or to excess emission in the central brightness depression of a black-hole shadow~\cite{Vincent:2020dij,Eichhorn:2022fcl,Carballo-Rubio:2022aed}. On the other hand, a common view point in quantum gravity suggests that quantum-gravitational effects should be negligible at the scale of the horizon of macroscopic black holes. Therefore, to determine the physical viability of the scale-identification based on local temperature further studies are required, elucidating the precise mechanism by which the RG-improvement produces an object in which horizon formation is avoided.
\\

In the UV regime, we assume a scale-dependence of the gravitational couplings dictated by their canonical mass dimension, which is consistent with the prediction of a non-Gaussian UV fixed-point according to asymptotic safety. The fixed-point scaling of the dimensionful couplings, combined with the scale-identification via the local blueshift factor, results in modified field equations in the UV which admit analytical solutions in the form of power laws for the $(0,0)$-component of the metric. One of these power-law solutions with the power-law exponent $\alpha=\sqrt{3}-1$ is singled out by our numerical solutions that smoothly connects to a Schwarzschild exterior. As a consequence, the metric at the origin becomes scale-invariant with respect to a rescaling of the radial coordinate. An intriguing question is whether a scale-invariant black hole core can be obtained in other approaches to quantum gravity using the language of critical phenomena, e.g., in loop quantum gravity, string theory, or in a holographic set-up.

The scalar curvatures of our solutions diverge at the origin and thereby signify the presence of a naked singularity. Moreover, the divergence in the Kretschmann scalar is stronger than for the Schwarzschild spacetime, cf.~Eq.~\eqref{eq: Kretschmann}. While it is unclear from a classical and quantum point of view, whether these types of singularities are physically relevant, these findings call for further investigations determining the physical consequences of the choice of scale-identification.

Let us emphasize, however, that in classical theories of gravity, curvature singularities and the physically more relevant question of geodesic completeness are not equivalent, see e.g.~\cite{Bejarano:2017fgz}. Spacetimes with integrable singularities~\cite{Lukash:2011hd,Lukash:2013ts} may exist, for which the presence of a curvature singularity does not lead to the geodesic incompleteness of the spacetime.
In the context of destructive interference of singular spacetimes in the path integral as a selection-mechanism between candidates for a microscopic action~\cite{Lehners:2019ibe,Borissova:2020knn}, it was pointed out~\cite{Giacchini:2021pmr} that for many regular black hole geometries with regular Riemann invariants (i.e., invariants built solely from the Riemann tensor), some of the derivative invariants (i.e., invariants built from covariant derivatives and the Riemann tensor) can still diverge. Therefore, even at the classical level, the characterization of geometries based on only local curvature invariants is most likely insufficient to characterize the physical viability of the spacetimes. In light of the previous remarks, determining the viability of the obtained quantum-corrected spacetimes as classical geometries, would require the identification and evaluation of physically meaningful observables.

At the quantum level, major challenges arise due to the obstructions in defining local observables (e.g.,~\cite{Donnelly:2015hta}) related to the diffeomorphism-invariance of gravity. In particular, the divergence of the local curvature may not be meaningful in the deep UV, where the dynamics is described by quantum gravity.

The existence of a naked time-like singularity within a strong-gravity region interior to the modified horizon structure in our model is reminiscent of the 2:2 hole solutions from quadratic gravity~\cite{Holdom:2002xy, Holdom:2016nek,Holdom:2019bdv} (see \cite{Yokokura:2022kmq} for a similar proposal). These are characterized by the leading terms $A(r)\propto r^2$ and $B(r)\propto r^2$ in a series expansion around the origin. Quadratic curvature terms modify the behavior in the UV compared to classical GR with zero cosmological constant, whereas in the case of an RG-improved Einstein-Hilbert action a qualitatively similar effect arises as a consequence of the running of the cosmological constant. For the 2:2 solutions, thermodynamic quantities of a finite-energy wavepacket remain bounded at the origin~\cite{Holdom:2016nek,Holdom:2019bdv,Holdom:2019ouz,Ren:2019afg}. A similar outcome may be conjectured for the fate of wavepackets passing through our scale-invariant cores, which is subject to a better understanding of the dissipative properties of thermal quantum systems in the deep UV.

It should be pointed out that the qualitative similarity with solutions from quadratic gravity is in line with the original motivation for an RG-improvement of couplings. As reviewed in Sec.~\ref{subsec: RG-improvement idea}, through an RG-improvement of the classical action, it may be possible to qualitatively study the effect of higher-order and nonlocal terms in the effective action which were not taken into account in the original truncation. In fact, recently, RG-improvement based on a choice of scale-identification from the decoupling mechanism has been applied to gravity~\cite{Borissova:2022mgd}. By identifying the decoupling scale associated with the effective action in the Einstein-Hilbert truncation and using this scale as an input for the RG-improvement, the dynamics of quantum-corrected black holes were studied in the stages of formation to evaporation. As a key result, the spacetimes derived in~\cite{Borissova:2022mgd} feature properties reminiscent of higher-derivative operators with non-local form factors.
	
Similarly, we observe a qualitative similarity between our solutions to an RG-improved Einstein-Hilbert action and
classical quadratic gravity solutions. Extending our analysis to an RG-improved quadratic gravity action with the same choice of cutoff, may reveal whether our result for the power-law exponent of the metric is only an artifact of the Einstein-Hilbert truncation of the action, or rather indicates a larger universality class. Yet another important question is the whether similar scale-invariant cores can also be found beyond spherical symmetry, e.g., for a stationary and axisymmetric spacetimes.

Let us reiterate the common expectation from effective field theory that quantum gravity should not induce corrections to macroscopic black holes. Such a perturbative viewpoint is based on the assumption that quantum gravity effects induce modifications to general relativity which are purely local. Nevertheless, non-local (and non-perturbative) contributions may arise and have been suggested in various approaches to quantum gravity~(e.g., \cite{Mathur:2005zp,Almheiri:2012rt,Ho:2021sbi,Borissova:2022clg}). Similarly, the question of what an infalling geodesic observer may see as they cross the horizon becomes ambiguous, as a quantum observer cannot be localized on a classical trajectory (e.g., see the Fuzzball complementarity proposal \cite{Avery:2012tf}).
\\

Finally, we stress that quantum-gravity phenomenology based on an RG-improve-ment of couplings should not be viewed as a first-principle derivation from asymptotic safety.
We emphasize that our results differ qualitatively from previous applications of RG-improvement by deviations from GR becoming apparent already at the classical horizon. 
The relevance of quantum gravity effects at the horizon is usually debated in relation to the information loss problem~\cite{Mathur:2005zp,Almheiri:2012rt}.

We attribute this qualitative difference to the different physical assumptions underlying the scale-identification of the RG-scale $k$ with an energy scale in the classical theory: RG improvement in the literature, cf.~\cite{Bonanno:2000ep} and~\cite{Platania:2019kyx} for a recent review, is based on a scale identification tied to local curvature. 
For astrophysically relevant black holes, the curvature at the event horizon is small compared to the Planck scale. As a result, modifications to astrophysically relevant black holes have been found to be small.
In contrast, the RG improvement in this paper is based on a scale-identification tied to local temperature. For any black hole, the local blueshifted temperature diverges at the horizon. As a result, modifications become significant at a Planck distance to the classical event horizon and result in a horizonless compact object. Whether the resulting object provides a realistic model for a quantum-corrected black-hole spacetime remains an open question and will ultimately have to be addressed by obtaining the full effective action and its respective asymptotically flat solutions.

The qualitative differences in the result of our RG-improvement hinge on the physical assumptions made in the respective scale-identification. One such fundamental assumption is the choice of input based on the classical solution. Differences compared to an RG-improvement with local curvature do not arise from considering only a subgroup of the full diffeomorphism group. Such a restriction is not strictly necessary. Qualitatively similar results, showing modifications at the classical horizon, can presumably be obtained from an RG-improvement at the level of the Schwarzschild solution with a horizon-detecting metric invariant given by the square of the covariant derivative of the Riemann tensor~\cite{Karlhede:1982fj}.

In principle, the RG-improvement could also be tied to a combination of scale identifications with local temperature and local curvature. In this context, the extremal limit of a charged or spinning black hole could be particularly interesting. As one approaches the extremal limit, the temperature of the classical black hole vanishes. Close to extremality one would thus expect a non-trivial competition between both scale identifications. We hope to return to this question in future work.

We conclude by emphasizing that, in a consistent treatment of asymptotic safety, physical quantum effects have to be calculated from the effective action, where all fluctuations have been integrated out.

\acknowledgments

We thank B.~Holdom and A.~Platania for invaluable discussions and comments. We also thank B.~Knorr for comments on the manuscript. A.~Held would like to thank the Perimeter Institute for Theoretical Physics for hospitality during the initial phase of this project.
The work leading to this publication was supported by the PRIME programme of the German Academic Exchange Service (DAAD) with funds from the German Federal Ministry of Education and Research (BMBF), by the Natural Sciences and Engineering Research Council of Canada, and by Perimeter Institute for Theoretical Physics. Research at Perimeter Institute is supported in part by the Government of Canada through the Department of Innovation, Science and Economic Development and by the Province of Ontario through the Ministry of Colleges and Universities.

\appendix

\section{Appendix}\label{app: differential equations}

Inserting a general cutoff-function $k^2(A(r))$ into the action~\eqref{eq: modified EH action} and varying with respect to the $(1,1)$-component of the metric, we can express the function $B(r)$ as a function of $A(r)$, 
\begin{equation}\label{eq: B rule}
B= \qty(-2 A G_k - 2 r G_k A' +  4 r A A' {G_k}' {k^2}' +  r^2 {A'}^2 {G_k}' {k^2}') \Big/\Big(2 A {G_k} (-1 + r^2 \Lambda_k)\Big)
\end{equation}
where $G_{k}=G\qty(k^2(A(r)))$, similar for $\Lambda_{k}$ and primes denote derivatives with respect to the argument. Subsequently,~\eqref{eq: B rule} can be inserted into the equation obtained by a variation of the action with respect to the function $A$. Specifying the scale-dependence of $G(k^2)$ and $\Lambda(k^2)$ according to~\eqref{eq: scale dependence} and the cutoff function according to~\eqref{eq: cutoff identification}, leads to a second-order nonlinear ordinary differential equation for $A(r)$,
\begin{align}\label{eq: general differential equation for A}
A'' = {} 
            & \Bigg(12 \chi \lambda_* T^4 \omega r A^4 (T^2 \omega + \chi A)  + 2 A^2  A' \Big(-9 \lambda_* T^8 \omega^3 r^2 \nonumber + A \Big(3 T^6 \omega^2 (-3 \chi \lambda_* r^2 + 2 \omega) \\
            & + 2 \chi A \Big(3 T^4 \omega (\chi \lambda_* r^2 + \omega) \nonumber + \chi A (T^2 \omega + \chi A)\Big)\Big)\Big) + 2 T^2 \omega r A  A'^2 \Big(9 \lambda_* T^6 \omega^2 r^2 \nonumber\\
            & +  A \Big(-3 T^4 \omega (3 \chi \lambda_* r^2 + 2 \omega) -  \chi A \Big(T^2 (-2 \chi \lambda_* r^2 + 3 \omega) + \chi A)\Big)\Big) \nonumber \\
            & + T^2 \omega r^2 A'^3 \Big(6 \lambda_* T^6 \omega^2 r^2 -  A \Big(3 T^4 \omega^2 + \chi A (3 T^2 \omega + 2\chi A)\Big)\Big)\Big) \Bigg) \Bigg/ \nonumber\\
            & \Bigg(2 r A^2 \Big(3 T^4 \omega^2+\chi^2 A^2\Big) \Big(\lambda_* T^4 \omega r^2 - A (T^2 \omega + \chi A)\Big)\Bigg)
\end{align}
For the exact theory in the fixed-point regime, we set $\chi= 0$ and find the power-law solutions for $A$ presented in Sec.~\ref{subsec: fixed-point analysis}. For the numerical integration we set $\chi = 1$, which reproduces the running of the dimensionful Newton coupling derived in~\cite{Bonanno:2000ep}. The latter interpolates between the classical constant Newton coupling in the IR and the power-law running in the UV regime. Additionally, we model the scale-dependence of the cosmological constant according to~\eqref{eq: scale dependence}, i.e.~a $k$-dependence consistent with the UV regime and at the same time imposing an effective zero cosmological constant in the IR. We make this choice as we are interested in asymptotically flat solutions. Demanding consistency of the numerical solutions with the exact solution to the fixed-point equations in the small-$r$ regime, shows that in the UV the power law $A\propto r^{\alpha}$ with exponent $\alpha = \sqrt{3}-1$ is realized. For this solution, inserting the fixed-point scaling of $G$ and $\Lambda$ into the solution rule for $B$,~\eqref{eq: B rule}, leads to $B=0$. Therefore, while the scaling law for $A$ is in the appropriate sense universal in a finite neighbourhood of the fixed-point at $r=0$, the leading term for $B$ at $r>0$ must be obtained by considering small $\chi >0$ corrections to $A$. We perturb the fixed-point equations by expanding~\eqref{eq: general differential equation for A} around $\chi=0$ to first order and examine which choices of $\alpha_1$ and $a_1$ in the first-order corrected function $A(r) = a_0 r^{\alpha} + \chi a_1r^{\alpha_1} $ solve the perturbed fixed-point equations to lowest order in $\chi$. This leads to $\alpha_1 = 2 \alpha$ and $a_1 ={a_0}^2(1 + \sqrt{3})/(2 T^2 \omega) $. Inserting the result for $A$ into~\eqref{eq: B rule}, keeping only the lowest-order terms in $\chi$, shows that $B$ receives $\chi^2$ corrections in a neighborhood of the fixed-point, $B(r)\propto {a_0}^2(7+2 \sqrt{3})/(2 T^4 \omega^2) r^{2\alpha}.$ The obtained $r$-dependence for $B$ correctly reproduces the numerical solution at small $r$.

\printbibliography

\end{document}